\documentstyle[12pt]{article}
\catcode`\@=11 
\newcommand{\half}{{\mbox{\small{$\frac{1}{2}$}}}}
\newcommand{\quarter}{{\mbox{\small{$\frac{1}{4}$}}}}
\def\be{\begin{equation}}
\def\ee{\end{equation}}
\def\beqar{\begin{eqnarray}}
\def\eeqar{\end{eqnarray}}
\def\barr{\begin{array}}
\def\earr{\end{array}}
\def\lsim{\:\raisebox{-0.5ex}{$\stackrel{\textstyle<}{\sim}$}\:}
\def\gsim{\:\raisebox{-0.5ex}{$\stackrel{\textstyle>}{\sim}$}\:}
\def\and{\qquad {\rm and } \qquad}
\def\ie{ {\it i.e.} }

\def\slp{p  \hspace{-1.0ex}/}
\def\slk{k  \hspace{-1.0ex}/}
\def\slq{q  \hspace{-1.0ex}/}

\def\etsl{$\not${\hbox{\kern -2.5pt $E_T$}} }

\def\lsim{\mathrel{\mathpalette\@versim<}}
\def\gsim{\mathrel{\mathpalette\@versim>}}
\def\@versim#1#2{\vcenter{\offinterlineskip
    \ialign{$\m@th#1\hfil##\hfil$\crcr#2\crcr\sim\crcr } }}
\begin{document}
\setcounter{page}{0}
\renewcommand{\thefootnote}{\fnsymbol{footnote}}
\begin{flushright}
CTP-TAMU-76/93\\
MRI-PHY-/12/93\\
LTH-326\\
\end{flushright}
\vglue 0.5cm
\begin{center}
{\Large\bf Two-Loop Neutrino Masses and the Solar Neutrino Problem\\}
\vglue 1cm
{\sc
Debajyoti Choudhury$^{(a)}$\footnote{debchou@iws166.mppmu.mpg.de,
  debchou@dmumpiwh.bitnet},
Raj Gandhi$^{(b),(c)}$\footnote{raj@tamphys.bitnet},
J.A. Gracey$^{(d)}$\footnote{jag@s-a.amtp.liv.ac.uk}
and Biswarup Mukhopadhyaya$^{(e)}$\footnote{biswarup@mri.ernet.in}
}
\vglue 0.4cm
{\em $^{(a)}$ Max-Planck-Inst. f\"{u}r Physik, Werner Heisenberg Inst.\\}
{\em F\"{o}hringer Ring 6, 80805 M\"{u}nchen, Germany\\}
{\em $^{(b)}$Center for Theoretical Physics, Department of Physics,
Texas A\&M University\\}
{\em College Station, TX 77843-4242, USA\\}
{\em $^{(c)}$Astroparticle Physics Group, Houston Advanced Research Center
(HARC)\\}
{\em The Woodlands, TX 77381, USA\\}
{\em $^{(d)}$ DAMTP, University of Liverpool,\\}
{\em P.O Box 147, Liverpool L69 3BX, U.K.\\}
{\em $^{(e)}$ Mehta Research Institute, 10 Kasturba Gandhi Marg,\\}
{\em Allahabad, 211002, India\\}
\vglue 1cm
{\tenrm \bf ABSTRACT}
\end{center}

\noindent The addition of $m$ singlet right-handed neutrinos to the Standard
Model leads to radiatively generated mass corrections for the $SU(2)_L $
doublet neutrinos. For those neutrinos which are massless at the tree level
after this addition, this implies a small mass generated at the two-loop
level via $W^{\pm}$ exchange. We calculate these mass corrections exactly
by obtaining an analytic form for the
general case of $n$ doublets and $m$ singlets. As a phenomenological
application, we consider the $m=1$ case and
 examine the masses and mixings of the doublet neutrinos
which arise as a result of the two-loop correction in the light of
 experimental data from two sources which may shed light on the question
of neutrino masses. These are
(a) the neutrino detectors reporting
a solar
neutrino deficit (and its resolution via Mikheyev-Smirnov-Wolfenstein matter
oscillations),
 and (b) the COBE satellite data on the non-zero
angular variations of the cosmic microwave background
temperature (and its possible implications for hot dark matter).
 Within the framework of the
extension
considered here, which leaves the gauge group structure of
the Standard Model intact, we  show that it is possible
for neutrinos to acquire  small masses naturally, with values
 which are compatible with
current theoretical bias and experimental data.
\vfill\eject
\setcounter{footnote}{0}
\renewcommand{\thefootnote}{\arabic{footnote}}
\renewcommand{\theequation}{\thesection.\arabic{equation}}
\newcommand{\sect} [1]{\section{#1} \setcounter{equation}{0}}
\setcounter{page}{1}
\pagestyle{plain}

\sect{Introduction}

It is fair to say that the problem of understanding the origin of
fermion masses is one of the most perplexing questions facing particle
physics today. The Standard model \cite{SM}
can reproduce the observed fermion masses via electroweak
symmetry breaking and the Higgs mechanism, but provides no explanation
for their values.
 When such an understanding is obtained, one of the issues
that it must clarify is the smallness of neutrino masses (if, indeed, neutrinos
are massive) relative to those of the other fermions. An attractive
explanation for this observed feature of the fermion mass spectrum is the
see-saw mechanism \cite{ssaw}.
It postulates the existence of right-handed neutrinos with masses
of the order of the next energy threshold and uses this in
combination with the Higgs mechanism
to generate
light (Majorana) neutrino masses via an effective dimension five operator.

Given our present ignorance of the origins of mass and
the lack of experimental pointers towards any particular mechanism, it
is important to keep an open mind on the smallness of neutrino
mass. In this paper, we explore, via detailed calculation, the issue of
radiatively generated neutrino masses, since this is also a natural way
in which  masses small compared to those of other fermions may be generated.

Any such effort needs to invoke physics beyond the Standard model. In view of
the extraordinary and  demonstrated robustness of the model to
experimental tests over the last twenty years, we have thought it reasonable
to make the simplest possible extension to the standard theory and study
its effect on neutrino masses via radiative corrections,
 \ie the addition of $m$  $SU(2)_L \bigotimes U(1)_Y$ singlet right-handed
neutrinos.
 {\it A priori}, there is no connection
between their number and that of the doublet neutrinos, hence the simplest
case corresponds to $m=1$, \ie the addition of
one right-handed singlet neutrino
to the standard model \cite{JA}.

The gauge group structure of the weak sector remains unchanged as a
consequence of this extension, but  majorana mass terms incorporating the
scale of new physics are now allowed. We do not speculate on their origin, but
only note that it would require invoking an additional global symmetry
(such as a conserved lepton number) to set these
 to zero. The doublet neutrinos acquire radiative (and, in some cases,
tree-level) masses due to the presence of the singlets, as we discuss below.
The radiative masses arise (via mixing) due to a two-loop mechanism
\cite{PT,BM} involving the exchange
of $W^{\pm}$ bosons.
In Sections II and the Appendix, we calculate, exactly and in analytic form,
 the two-loop masses
 accquired by the initially massless doublet neutrinos.
Our calculation is general and valid for any number of doublet
and singlet neutrinos, but in order to obtain phenomenologically useful
information, we focus, in Section III, on the $m=1$  case.
Even this simplest extension
of the Standard model
introduces four new parameters into the theory. On the issue of neutrino
masses, it is non-accelerator
experiments that provide information on the cutting edge. Hence we have chosen
to examine
 the results for the $m=1$  case
  in the light of (a) the MSW {\cite{MSW}}
solution to the solar neutrino deficit seen by the  Kamiokande \cite{KII},
GALLEX \cite{GAL}, SAGE \cite{SA} and  Homestake \cite{Cl} neutrino detectors
and (b) the
implications for hot dark matter (neutrinos) from the recent COBE observations
on the anisotropy of the microwave background
\cite{COB}.
Invoking this experimental information restricts
the parameter space and consequently, in the context
considered in this paper, permits a handle on
the range of the mass scale
characterizing physics beyond the standard model. We show that doublet neutrino
masses compatible with both (a) and (b) above can result naturally from
such physics at the several hundred GeV scale.

\sect{Radiative Generation of Neutrino Masses }

In this section
we give a description of an exact general procedure for calculating
two-loop neutrino masses
 applicable
to any extension of the Standard model which incorporates singlet right-handed
neutrinos. (We remark below on the reason why a one-loop mass
does not arise in the sitiuation considered here, where {\it only}
right-handed handed neutrinos are added to the existing particle spectrum. )
After setting up the generic integral that needs to be
calculated we describe the procedure for evaluating it exactly in the Appendix.

The lepton sector of the extension considered here has, in general,
$n \:(\geq 3)$ doublet fields $[\nu'_{iL} \;l_{iL} ]^T $
 and $m$ singlet fields $(\nu'_{A L})^c = (\nu_A^{\prime c})_R
$. ( Here $i=1....n$, $A=1....m$ and $\nu^c \equiv C \bar\nu^T$ is the charge
conjugate spinor.)
In addition, one has the charged lepton $SU(2)_L$ singlet fields $l_{iR}$.
The primes on the neutrino fields denote weak
eigenstates as opposed to physical particle states.
Without any loss of generality,  we have  assumed that the
weak eigenstates $l_{i}$ are the same as the corresponding mass eigenstates,
\ie the charged lepton mass matrix is diagonal.

As noted in the Introduction,
in addition to the Dirac mass terms,
the most general Lagrangian consistent with
the gauge symmetry of the Standard model
 also contains possible Majorana mass terms for neutrinos
 of the form
$m_{AB} \overline{\left(\nu'_{A L}\right)^c} \nu'_{B L}$. In the minimal
model under consideration here,
such terms must
be bare mass terms, but in a more involved model they could arise,
for instance,
due to the  vacuum expectation value of a
singlet higgs.  To facilitate
discussion, we combine all
the left handed neutrinos into a $(n+m)$--dimensional vector in the flavour
space denoting it by $\nu'_{\alpha L}$.
The most general mass term is thus given by
\be
{\cal L}_{\rm m} = \sum_{i = 1}^n \mu_i \bar{l}_{iL} l_{iR}
    + \sum_{\alpha,\beta = 1}^{n+m}
          \overline{\left(\nu'_{\alpha L}\right)^c}
                   {\cal M}_{\alpha \beta} \nu'_{\beta L} + h.c
      \label{massinwk}
\ee
Here $\cal M$ is a complex
symmetric \footnote{That $\cal M$ has to be symmetric is evident from the
charge conjugation property of fermion bilinears.}
$(n+m) \times (n+m)$ matrix of the form
\be \displaystyle
{\cal M} = \pmatrix{0_{n \times n} & D_{n \times m} \cr
                        D^T_{m \times n} & M_{m \times m}\cr}
  \label{massmatrix}
\ee
with $D$ and $M$ denoting the Dirac and the Majorana  mass terms respectively.
The first block is identically zero in the absence of a non--trivial vacuum
expectation value for a $SU(2)_L$--triplet
higgs field. (This restriction is imposed not only by our philosophy of
minimal extension, but more importantly, by $m_W/m_Z$ --- the observed ratio
of the
gauge boson masses.) $\cal M$ can be diagonalized by
a biunitary transformation of the form
\be
V^T {\cal M} V =  \widehat{{\cal M}} = {\rm diagonal}(m_\alpha)
       \label{biunitary}
\ee
The mass eigenstates ( $\nu_\alpha$) are then easily identified to be
\be
\nu_L = V^\dagger \nu'_L
     \label{basischange}
\ee
The relevant piece of the  weak Lagrangian is then given by
\be
{\cal L}_{\rm wk} = J_\mu W^\mu
    \label{lagrangian}
\ee
where
\be
\barr{rcl}
J_\mu^+ & = & \displaystyle \frac{ig}{\sqrt{2}}
             \sum_{i = 1}^n \bar{l}_{i} \gamma_\mu P_L \nu'_i
 =  \displaystyle \frac{ig}{\sqrt{2}}
             \sum_{i = 1}^n \sum_{\alpha =1}^{n+m} K_{i\alpha}
             \bar{l}_{i} \gamma_\mu P_L \nu_i
     \\[2ex]
J_\mu^3 & = & \displaystyle \frac{ig}{2 c_W}
             \sum_{i = 1}^n \left(
                 \bar{l}_{i} \gamma_\mu P_L \l_i
                 + \bar{\nu'}_{i} \gamma_\mu P_L \nu'_i \right) \\[2ex]
& = & \displaystyle \frac{ig}{2 c_W} \left[
             \sum_{i = 1}^n  \bar{l}_{i} \gamma_\mu P_L \l_i
                 + \sum_{\alpha, \beta = 1}^{n+m}
                  (K^\dagger K)_{\alpha \beta}
                  \bar{\nu}_{\alpha} \gamma_\mu P_L \nu_\beta \right]
     \label{currents}
\earr
\ee
Here $c_W \equiv \cos \theta_W$, where $\theta_W$ is the Weinberg angle,
$g = e/\sin\theta_W$,
 $P_L \equiv (1 - \gamma_5)/2$ and
\be \displaystyle
 K = \pmatrix{I_{n \times n} & 0 \cr
              0 & 0\cr} V
    \label{CKM}
\ee
is the $(n+m)$--dimensional analog of the quark sector
Cabibbo--Kobayashi--Maskawa matrix. Note that though
$K K^\dagger = {\rm diag}(I_{n \times n},0)$,
$(K^\dagger K)_{\alpha \beta} \neq \delta_{\alpha \beta}$. Thus we do indeed
have flavour changing neutral currents (FCNC) in the neutrino sector.

Having set up the general formalism, let
 us now concentrate on the case where $n > m$.  There exist then $n - m$
neutrinos that are strictly massless at the tree level. We now
calculate the changes to such a spectrum accruing from
quantum corrections.

 Before proceeding,
in view of the  fact there exist massive neutrinos and also FCNC's in the
neutrino sector, it is appropriate at this point to remark on the possibility
of  one--loop graphs with a
$Z$ or a Higgs exchange  introducing a non--trivial correction
to the neutrino mass matrix. However, it can be easily seen that it is
possible to rotate the neutrino states such that only $m$ of them have
Yukawa couplings to the Higgs. Thus, only those doublet states that are
massive at tree level obtain a Higgs induced mass at the one-loop level.
In addition, since the flavor-changing $Z$ couplings have the same mixing
parameters as the flavor-changing Yukawa couplings, the one-loop
$Z$ exchange diagrams do not contribute to the masses of the $n-m$
neutrinos which are massless at the tree level. This reasoning
applies at all orders to any diagram where all virtual particles are
neutral.
Hence
the relevant diagram to compute is that given in Fig.(1).

We shall work in the weak interaction basis for the external neutrinos and the
mass basis for all the virtual particles. Furthermore, we shall concentrate
only on the first $n \times n$ block of ${\cal M}$, \ie on the
generation of Majorana mass terms for the doublet neutrinos.
In the unitary gauge,  the correction to the neutrino propagator
is then given by
\be
\barr{rcll}
i \Sigma^{(2)}_{ij} (p) &= &\displaystyle
                 \left(\frac{ig}{\sqrt{2}} \right)^4\;
                    \sum_{\alpha = 1}^{n + m}
                   \int
                 & \displaystyle \!\! \! \!
                   \frac{d^4 k}{(2 \pi)^4} \frac{d^4 q}{(2 \pi)^4}
                   \gamma_\mu P_R \frac{i}{\slp + \slq - \mu_i}
                   K^\dagger_{\alpha i} \gamma_\nu P_R
                        \\[1.5ex]
             &&& \hspace*{2ex} \displaystyle
   \frac{i}{\slp + \slq + \slk - m_\alpha}
                   K^\dagger_{\alpha j}
                   \gamma_\sigma P_L \frac{i}{\slp + \slk - \mu_j}
                   \gamma_\lambda P_L
                       \\[1.5ex]
             &&& \hspace*{2ex} \displaystyle
       \frac{ -i \left(g^{\mu \sigma} - q^\mu q^\sigma /m_W^2 \right)}
            {q^2 - m_W^2}\:
       \frac{ -i \left(g^{\nu \lambda} - k^\nu k^\lambda /m_W^2\right)}
            {k^2 - m_W^2}
\earr
     \label{2loopmass-1}
\ee
The mass correction is of course given by
${\cal M}^{(2)}_{ij} = \Sigma^{(2)}_{ij}(p = 0) $, and
after some algebra this leads to
\be
\barr{rcll}
{\cal M}^{(2)}_{ij} & = & \displaystyle
             \frac{g^4}{4}
                   \sum_{\alpha = 1}^{n + m} m_\alpha
                   K^\dagger_{\alpha j} K^\dagger_{\alpha i}
                   \int
                 & \displaystyle \! \! \! \!
                   \frac{d^4 k}{(2 \pi)^4} \frac{d^4 q}{(2 \pi)^4}
                     \frac{(k + q)^2\,k \cdot q }{ {\cal D}_{ij;\alpha}}
\\[2ex]
                 &&& \displaystyle
                   \left[ \left(4 + \frac{k^2 q^2}{m_W^4} \right)
                           - 4\,\frac{q^2 + k^2}{m_W^2} \right]
\earr
         \label{mass}
\ee
where
\be
{\cal D}_{ij;\alpha} = (k + q)^2 \left\{(k + q)^2 - m_\alpha^2 \right\}
                   (q^2 - \mu_i^2) (q^2 - m_W^2) (k^2 - m_W^2) (k^2 - \mu_j^2)
   \label{denom}
\ee
We see that the mass corrections would be identically zero if $m_\alpha =0,
\; \forall \alpha$. This ought to be so as any mass renormalization must be
proportional to the bare mass terms. The integral above has a naive degree of
 divergence of 4. However, note that
\be
\sum_{\alpha = 1}^{n + m} m_\alpha
                   K^\dagger_{\alpha j} K^\dagger_{\alpha i}
       = {\cal M}_{i j} = 0
\ee
and hence
\be
\sum_{\alpha = 1}^{n + m} m_\alpha K^\dagger_{\alpha i} K^\dagger_{\alpha j}
                  \frac{(k+q)^2}{(k+q)^2 - m_\alpha^2}
= \sum_{\alpha = 1}^{n + m}
                  \frac{ K^\dagger_{\alpha i} K^\dagger_{\alpha j} m_\alpha^3}
                       {(k+q)^2 - m_\alpha^2}
   \label{GIM-1}
\ee
This clearly is analogous to the GIM mechanism in the quark sector. Even on
substitution of eqn(\ref{GIM-1}) in eqn(\ref{mass}), the integral in the latter
is still formally divergent. Notice, however, that this is but
an artifact of the
unitary gauge and is not a real divergence \cite{div}.
In fact, by invoking identities
similar to eqn(\ref{GIM-1}) or equivalently, by working in the Feynman gauge,
one obtains \footnote{This has often been cited in the literature as a
GIM-like cancellation, but in our view the two are quite different.}
\be
{\cal M}^{(2)}_{ij}  =  \displaystyle
                   g^4
                   \sum_{\alpha = 1}^{n + m} m_\alpha^3
                   K^\dagger_{\alpha j} K^\dagger_{\alpha i}
                   \left[ 4 + 4 \frac{\mu_i^2 + \mu_j^2}{m_W^2}
                       + \frac{\mu_i^2 \mu_j^2}{m_W^4} \right]
                    \Lambda(\mu_i^2, m_W^2, m_\alpha^2,0,\mu_l^2, m_W^2 )
         \label{mass-final}
\ee
where
\begin{eqnarray}
&&  \Lambda(m_1^2,m_2^2,m_3^2,m_4^2,m_5^2,m_6^2) \equiv \nonumber \\
&&\!\!\!\!\!  \int \frac{d^4 k \;d^4 q\; k \cdot q}
               {   (q^2 + m_1^2) (q^2 + m_2^2)
                  \left\{(k + q)^2 + m_3^2\right\}
                   \left\{(k + q)^2 + m_4^2 \right\}
                   (k^2 + m_5^2) (k^2 + m_6^2)
               } \nonumber \\
    \label{Lambda-defn}
\end{eqnarray}
is an Euclidean integral evaluated in the Appendix.

The expression in eqn(\ref{mass-final}) thus represents the Majorana mass
generated for the doublet neutrino at the two--loop level. In operator
language, it arises from  terms of the form
\be
   \overline{(L_{iL})^c} L_{jL } \phi \phi S
       \label{operator}
\ee
where $L_{iL}$ represent the doublet lepton fields, $\phi$ is the usual
higgs field and $S$ represents the lepton number violating operator (whether
a singlet higgs or a bare mass term). We note that this five dimensional
effective operator for the radiative masses is the same as that for the
conventional see-saw mechanism. The difference between the two resides
in the scale of mass generation. Two-loop radiative masses compatible with the
solar and COBE data can arise from right-handed neutrinos at the several
hundred GeV scale, as we show below, whereas the see-saw mechanism generates
similar valued masses via heavy neutrinos at the  grand unified scale.

 We also note that though the corrections ostensibly are
proportional to $m_\alpha^3$ (eqn.(\ref{mass-final})),
 the actual dependence is linear (apart from logarithmic corrections) due to
suppressions hidden in $\Lambda$. As $m_\alpha$ becomes larger
and terms of the order of $(\mu_i/m_\alpha)^2$ become negligible, the
correction goes as $\Sigma_{\alpha} K^{\dagger}_{\alpha i} K^{\dagger}_{
\alpha j} m_{\alpha}$, which is simply the $(ij)^{\rm th}$
 element of the tree-level
mass matrix, and hence zero for the cases of interest here.

Finally, we remark
 that a complex ${\cal M}$ in eqn.(\ref{massinwk}) obviously leads
to a complex digonalizing matrix $V$ and hence possibly to $CP$--violating
processes. However, since there is no evidence as yet of any such
non-conservation in the leptonic sector, we have, in the interests of
 simplicity,
  chosen to perform all numerical calculations assuming a
real neutrino mass matrix.

\section{Application: The Solar Neutrino Deficit and COBE Data}

In order to make a connection to experiment and phenomenology, we
now
specialize to the $n=3$ and $m=1$ case and examine the two loop
mass corrections in the context of (a) the MSW solution \cite{MSW} to the solar
neutrino deficit reported by various detectors \cite{KII,GAL,SA,Cl} and
(b) recent COBE \cite{COB} data and its implications for neutrinos as dark
matter.

The solar deficit is the only long-standing possible evidence for physics
beyond the Standard model, and the MSW mechanism is its most popular
resolution. In its essence, the mechanism requires neutrinos to be
massive (and non-degenerate), allowing the interaction eigenstate
$\nu_e$ (assumed to comprise predominantly of the lightest mass eigenstate)
to oscillate to $\nu_{\mu}$ or $\nu_{\tau}$ due to the difference in the
forward scattering potential seen by the two states in their passage through
solar matter. It thus identifies a range of vacuum mixing angle and mass
squared difference  values which are compatible with the deficit observed
by the various detectors.
 Figure 2, excluding curves labelled (a), (b) and (c),
is taken from Ref. \cite{Hata} and shows the familiar two-flavor mixing
MSW solution space, where $\theta$ is the Cabibbo mixing angle and $\Delta m^2$
is the difference of the squares of the two neutrino masses, which, in the
present
context, are acquired at the two-loop level.

 COBE data
on the anisotropy of the microwave background, while not
making a definitive statement on the nature of dark matter, seem to suggest
that
it may have both hot and cold components, with the former
being a neutrino (since it is the only known hot dark matter candidate)
with a mass of $\approx 10 $ eV.

 We use both of the above considerations to restrict
the rather large parameter space available to us.

In the scenario with one additional singlet, we have two  massive and two
massless neutrinos at tree-level. The two massive ones acquire both
one-loop and two-loop corrections, which we neglect, and the massless
states acquire small masses at the two-loop level. The  two tree-level masses
 and all the radiative corrections are expressible in terms of four input mass
parameters for the matrix ${\cal M}$.
 For various plausible
 (fixed)  values
of $m_{\alpha}$, (the singlet mass, signifying the scale of new physics)
and the added constraint that the  other neutrino with a tree level mass
lie in the $10 $ eV range, we obtain a one parameter set of curves (see
Fig. $2$) which denotes the intersection of the "two-loop space" with
the MSW solution space. Note that restricting ourslves to
the two dimensional MSW space imposes
a third constraint, \ie that the $\nu_e$ mixes predominantly with
only one other state.
Curve (a) in Figure $2$ corresponds to a singlet mass
of $100 $ GeV and a $\nu_{\tau}$ mass of $\approx 8.6
$ eV. The two-loop masses  and mixings of $\nu_{e}$ and $\nu_{\mu}$
are then such that they span the MSW space as shown. Curve (b) corresponds
to a singlet mass of $ 400 $ GeV and a $ \nu_{\mu}$ mass of $\approx 7 $ eV.
$\nu_{\tau}$ and $\nu_{e}$ then acquire radiative masses
 and mixings that span the
solution space as shown. For sin$^22\theta$ greater than $\approx 3 \times
10^{-1}$,
$\nu_{\tau}$ becomes lighter than $\nu_{e}$, and MSW oscillations occur
between anti-neutrino rather than neutrino states, and are thus not relevant.
We note that (b) passes through the (small-angle, non-adiabatic)
 MSW region that is compatible with
all detectors and also represents a value of $m_{\nu_{\mu}}$ ($7$ eV)
that provides a very good fit to COBE data in the context of
a hot plus cold dark matter scenario.
 Finally, curve (c) represents a singlet mass of $1 $ TeV and a $\nu_{\mu}$
mass of $\approx 9.8 $ eV, and terminates where it does because for larger
mixing angles the $\nu_{e}$ becomes heavier than the $\nu_{\tau}$.
Note that the determination
of which flavor the $\nu_e$ oscillates to is made by examining the
mixing (diagonalizing) matrix of the full (\ie tree + loop ) mass matrix.
 A (reasonable) assumption
built into the results is that $\nu_e$ is the lightest state.

We stress that these curves represent a phenomenological
exercise more than anything else to demonstrate that our calculations can
 make connection with
experiment when the full parameter
space, which is quite large, is constrained by imposing physically and
empirically well-motivated restrictions.

We note that the singlet mass values chosen by us ($100$ GeV,
$400$ GeV and $1$ TeV) are
 not in conflict with accelerator \cite{RPP} or cosmological \cite{OL}
bounds on these particles.

Finally, we remark that a disparity between
the mass scales of the $\nu_{e}, \nu_{\mu} (\nu_{\tau})$ and that of
the $\nu_{\tau} (\nu_{\mu})$
seems to be required if we take both the solar and COBE implications
for neutrino masses seriously. In the simple model under consideration
here, such a disparity arises naturally since the neutrino which contributes
to dark matter has a tree level mass while the other two have loop masses.

\section{\bf Conclusions}

We have explicitly obtained an analytic
form for the radiative two-loop masses acquired by doublet neutrinos
in models where right-handed singlets are present. We have made an effort
to keep our calculation general and the expression for the mass correction
that we obtain may have applications in other models with right-handed
neutrinos. We have calculated
these masses (for the one singlet case) in the light of
experimental data from solar neutrino detectors
and from  COBE, within the confines of
the MSW solution to the solar deficit. By doing so we have made an effort
 to demonstrate
that intermediate scale physics  (\ie physics at $\leq 1$  TeV) can lead, in
a simple way, to naturally small masses for neutrinos which have physically
meaningful values, without requiring drastic changes in
the presently known particle spectrum
 or gauge group structure.

\appendix

\sect{\bf Appendix: Evaluation of $\Lambda_{123456}$}
In this section we discuss the exact evaluation of the fundamental finite
two loop four dimensional integral underlying the mechanism. As a first step,
though, we consider the more general two loop Euclidean space integral,
$\Lambda_{123456}$, defined by
\begin{equation}
\Lambda_{123456} =
\int_{p \, q} \frac{p \cdot q}{(p^2+m^2_1)(p^2+m^2_2)
((p+q)^2+m^2_3)((p+q)^2+m^2_4)(q^2+m^2_5)(q^2+m^2_6)} \nonumber \\
\end{equation}
which we will evaluate analytically and then specialize to the case we are
concerned with. For reasons which we explain below we choose to calculate eqn
(A.1) in $d$-dimensions where
\begin{equation}
\int_k ~=~ \frac{\mu^{4-d}}{(2\pi)^{d}} \int \, d^dk
\end{equation}
and $\mu$ is an arbitrary mass parameter introduced to ensure the coupling
constant remains dimensionless in our $d$-dimensional manipulations. The
subscripts on $\Lambda_{123456}$ correspond to the masses $m^2_i$ of the
integral and we note that the function has certain obvious symmetries,
$\Lambda_{123456}$ $=$ $\Lambda_{213456}$ $=$ $\Lambda_{563412}$, which ought
to be preserved in the final expression. The strategy to evaluate eqn (A.1)
is to use partial fractions to obtain a sum of $2$-loop integrals with three
propagators and then to substitute for the value of each of these
sub-integrals, which have been considered by other authors in different
contexts before, \cite{J0,J1,J2,J3}. For instance, if we define
\begin{equation}
J_{ijk} ~=~ \int_p \int_q \frac{p \cdot
q}{(p^2+m^2_i)(q^2+m^2_j)((p+q)^2+m^2_k)}
\end{equation}
then eqn (A.1) is built out of a sum of eight such integrals where its only
symmetry is $J_{ijk}$ $=$ $J_{jik}$. Rewriting the numerator of eqn (A.3) one
finds
\begin{equation}
J_{ijk} ~=~ \half [I_iI_j - I_jI_k - I_kI_i - (m^2_k-m^2_i-m^2_j)I_{ijk}]
\end{equation}
where
\begin{eqnarray}
I_i &=& \int_p \frac{1}{(p^2+m^2_i)} \\
I_{ijk} &=& \int_p \int_q \frac{1}{(p^2+m^2_i)((p+q)^2+m^2_j)(q^2+m^2_k)}
\end{eqnarray}
and the latter function is totally symmetric, corresponding to a two loop
vacuum
graph (ie zero external momentum). The integral $I_{ijk}$ has been considered
in \cite{J0,J1} and a single integral representation of it exists,
\cite{J2,J3,J4}. For our purposes, however, we have chosen to use the elegant
formula given in \cite{J4} since it is explictly symmetric in the masses.
Although $\Lambda_{123456}$ is itself ultraviolet finite the sub-integrals,
eqns (A.3) and (A.4), are divergent and therefore require regularization. In
\cite{J3,J4} dimensional regularization was introduced to control these
infinities, which is why we choose to calculate eqn (A.1) in $d$-dimensions, so
that $I_{ijk}$ involves double and simple poles in $\epsilon$ where $d$ $=$ $4$
$-$ $2\epsilon$. Therefore in the final result these must cancel for all
$m^2_i$. As a first step, it is trivial to observe that in the partial fraction
decomposition of eqn (A.1) the $I_iI_j$ type terms, which are also divergent,
formally cancel to leave only the $I_{ijk}$ terms. To proceed we recall the
important properties of $I_{ijk}$ which have been discussed in more detail in
\cite{J4}. In $d$-dimensions the exact value, for arbitrary $(\mbox{mass})^2$,
$x$, $y$ and $z$, is
\begin{equation}
I(x,y,z) ~=~ I(2a,0,0) + \Gamma^\prime [ F(\half c-y) + F(\half c-z)
- F(x-\half c) ]
\end{equation}
where
\begin{eqnarray}
I_{ijk} &=& I(m^2_i,m^2_j,m^2_k) \nonumber \\
\Gamma^\prime &=& \frac{(\mu^2)^{4-d}}{(4\pi)^d} \Gamma(2-\half d)
\Gamma(1 - \half d) \nonumber \\
a &=& \half [ x^2 + y^2 + z^2 - 2xy - 2yz - 2zx]^{1/2} \nonumber \\
c &=& x + y + z
\end{eqnarray}
and
\begin{equation}
F(w) ~=~ \int_a^w ds \, \frac{1}{(s^2-a^2)^{(4-d)/2}}
\end{equation}
The result (A.7) is valid in the region of $(x,y,z)$ space where $a^2$
$\geq$ $0$. For the case when $a^2$ $<$ $0$, then the solution is, with
$b^2$ $=$ $-$ $a^2$,
\begin{eqnarray}
I(x,y,z) &=& - \, I(2b,0,0) \sin (\half\pi d) \nonumber \\
&+& \Gamma^\prime [ G(\half c-x) + G(\half c-y) + G(\half c-z) ]
\end{eqnarray}
where
\begin{equation}
G(w) ~=~ \int_0^w ds \, \frac{1}{(s^2+b^2)^{(4-d)/2}}
\end{equation}
and, for example,
\begin{equation}
I(x,0,0) ~=~ \frac{\Gamma(2-\half d)\Gamma(3-d)\Gamma^2(\half d-1) x^{d-3}}
{(4\pi)^d\Gamma(\half d) (\mu^2)^{d-4}}
\end{equation}
which is clearly singular in four dimensions. To obtain the finite part of
$\Lambda_{123456}$ each part of $I(x,y,z)$ needs to be expanded in powers of
$\epsilon$ to the $O(1)$ term and the poles in $\epsilon$ cancelled. The
non-trivial part of this exercise is the $\epsilon$-expansion of the
$F(w)$ and $G(w)$ integrals. These have been given in \cite{J4} and we record
that to the $\epsilon$-finite term,
\begin{eqnarray}
(4\pi)^4 I(x,y,z) &=& - \, \frac{c}{2\epsilon^2} - \frac{1}{\epsilon} \left[
\frac{3c}{2} - L_1 \right] - \half [ L_2 - 6L_1 + \xi(x,y,z) \nonumber \\
&&+~ c(7+\zeta(2)) + (y+z-x)\overline{\ln}y \overline{\ln}z \nonumber \\
&&+~  (z+x-y)\overline{\ln}z \overline{\ln}x
+ (y+x-z)\overline{\ln}y\overline{\ln}x ]
\end{eqnarray}
where $\zeta(n)$ is the Riemann zeta function, $L_i$ $=$ $x\overline{\ln}^i x$
$+$ $y\overline{\ln}^i y$ $+$ $z\overline{\ln}^i z$, $\overline{\ln}x$ $=$
$\ln(x/\hat{\mu}^2)$, $\hat{\mu}^2$ $=$ $4\pi e^{-\gamma}\mu^2$ and $\gamma$
is Euler's constant, and for $a^2$ $>$ $0$,
\begin{equation}
\xi(x,y,z) ~=~ 8a [ M(\phi_z) + M(\phi_y) - M(-\phi_x)]
\end{equation}
where
\begin{equation}
M(t) ~=~ - \, \int_0^t d \phi \, \ln \sinh \phi
\end{equation}
and the angles $\phi_x$ are defined by
\begin{equation}
\phi_x ~=~ \coth^{-1} \left[ \frac{\half c - x}{a} \right]
\end{equation}
For $a^2$ $<$ $0$, then
\begin{equation}
\xi(x,y,z) ~=~ 8b[L(\theta_x)+L(\theta_y)+L(\theta_z) - \half \pi \ln 2]
\end{equation}
where the $\theta_x$ angles are given by
\begin{equation}
\theta_x ~=~ \tan^{-1} \left[ \frac{\half c - x}{b} \right]
\end{equation}
and $L(t)$ is the Lobachevskij function,
\begin{equation}
L(t) ~=~ - \, \int_0^t d \theta \, \ln \cos \theta
\end{equation}
Equation (A.17) can also be rewritten as
\begin{equation}
\xi(x,y,z) ~=~ 8b[\tilde{L}(\theta_z)+\tilde{L}(\theta_y)-\tilde{L}(-\theta_x)]
\end{equation}
where $\tilde{L}(t)$ $=$ $\int_t^{\pi/2} d\theta \, \ln \cos \theta$ in order
to make the obvious analytic continuation across $a^2$ $=$ $0$ more apparent.
It is worth noting that essentially eqn (A.1) has been reduced to a single
simple function, eqn (A.19), whose properties are well known. We have used the
following identities in order to write an efficient programme to calculate
$\Lambda_{123456}$ for a range of physical mass values. For instance,
\cite{J5},
\begin{eqnarray}
L(t) &=& - \, L(-t) \quad\quad\quad\quad\quad\quad\quad
\quad\quad\quad\quad\quad ~~~~~\mbox{for} ~
- \, \half\pi\leq t \leq \half \pi \nonumber \\
L(t) &=& L(\half\pi - t) + (t - \quarter \pi) \ln 2 - \half
L(\half\pi-2t) ~~~~\mbox{for} ~ 0 \leq t \leq \quarter \pi \nonumber \\
L(t) &=& \pm L(\pi\pm t) \mp \pi \ln 2
\end{eqnarray}
Therefore, when the argument of the Lobachevskij function is known, the
identities of eqn (A.21) mean that one need only write a routine to evaluate
$L(t)$ numerically in the range $[0,\half\pi)$. For example, if $0$ $\leq$
$\lambda$ $<$ $2\pi$ then for any integer $n$
\begin{equation}
L(2\pi n + \lambda) ~=~ 2\pi n \ln 2 + L(\lambda)
\end{equation}
and so on.

Returning to the partial fraction form of $\Lambda_{123456}$ with the result
for $I_{ijk}$, the $c$ and $L_i$ terms of the $\epsilon$ expansion cancel in
the final expression and we can therefore take the limit back to four
dimensions, $\epsilon$ $\rightarrow$ $0$. Consequently, we end up with the
following analytic expression:
\begin{eqnarray}
\Lambda_{123456} &=& - \, \frac{1}{4(4\pi)^4(m^2_1-m^2_2)(m^2_3-m^2_4)
(m^2_5-m^2_6)} \nonumber \\
&\times& \left[ (m^2_3-m^2_1-m^2_5) \left[ \xi_{135}
- m^2_1 \ln \left(\frac{m^2_1}{m^2_3}
\right) \ln \left(\frac{m^2_1}{m^2_5}\right) \right. \right. \nonumber \\
&&- \left. \left. m^2_3  \ln \left(\frac{m^2_3}{m^2_1}\right)
\ln \left(\frac{m^2_3}{m^2_5}\right)
- m^2_5 \ln \left(\frac{m^2_5}{m^2_1}\right)
\ln \left(\frac{m^2_5}{m^2_3}\right) \right] \right. \nonumber \\
&-& \left. (m^2_3-m^2_1-m^2_6)\left[ \xi_{136}
- m^2_1 \ln \left(\frac{m^2_1}{m^2_3}\right)
\ln \left(\frac{m^2_1}{m^2_6}\right) \right. \right. \nonumber \\
&&- \left. \left. m^2_3 \ln \left(\frac{m^2_3}{m^2_1}\right)
\ln \left(\frac{m^2_3}{m^2_6}\right)
- m^2_6 \ln \left(\frac{m^2_6}{m^2_1}\right)
\ln \left(\frac{m^2_6}{m^2_3}\right) \right] \right. \nonumber \\
&-& \left. (m^2_4-m^2_1-m^2_5) \left[ \xi_{145}
- m^2_1 \ln \left(\frac{m^2_1}{m^2_4}\right)
\ln \left( \frac{m^2_1}{m^2_5}\right) \right. \right. \nonumber \\
&&- \left. \left. m^2_4 \ln \left(\frac{m^2_4}{m^2_1}\right)
\ln \left(\frac{m^2_4}{m^2_5}\right)
- m^2_5 \ln\left(\frac{m^2_5}{m^2_1}\right)\ln\left(\frac{m^2_5}{m^2_4}\right)
\right] \right. \nonumber \\
&+& \left. (m^2_4-m^2_1-m^2_6)\left[ \xi_{146}
- m^2_1 \ln \left(\frac{m^2_1}{m^2_4}\right)
\ln \left(\frac{m^2_1}{m^2_6}\right) \right. \right. \nonumber \\
&&- \left. \left. m^2_4 \ln \left(\frac{m^2_4}{m^2_1}\right)
\ln \left(\frac{m^2_4}{m^2_6}\right)
- m^2_6 \ln \left(\frac{m^2_6}{m^2_1}\right)
\ln \left(\frac{m^2_6}{m^2_4}\right) \right] \right. \nonumber \\
&-& \left. (m^2_3-m^2_2-m^2_5)\left[ \xi_{235}
- m^2_2 \ln \left(\frac{m^2_2}{m^2_3}\right)
\ln \left(\frac{m^2_2}{m^2_5}\right) \right. \right. \nonumber \\
&&- \left. \left. m^2_3 \ln \left(\frac{m^2_3}{m^2_2}\right)
\ln \left(\frac{m^2_3}{m^2_5}\right)
+ m^2_5 \ln \left(\frac{m^2_5}{m^2_2}\right)
\ln \left(\frac{m^2_5}{m^2_3}\right) \right] \right. \nonumber \\
&+& \left.(m^2_3-m^2_2-m^2_6) \left[ \xi_{236}
- m^2_2 \ln \left(\frac{m^2_2}{m^2_3}\right)
\ln \left(\frac{m^2_2}{m^2_6}\right) \right. \right. \nonumber \\
&&- \left. \left. m^2_3 \ln \left(\frac{m^2_3}{m^2_2}\right)
\ln \left(\frac{m^2_3}{m^2_6}\right)
- m^2_6 \ln \left(\frac{m^2_6}{m^2_2}\right)
\ln \left(\frac{m^2_6}{m^2_3}\right) \right] \right. \nonumber \\
&+& \left.(m^2_4-m^2_2-m^2_5)\left[ \xi_{245}
- m^2_2 \ln \left(\frac{m^2_2}{m^2_4}\right)
\ln \left(\frac{m^2_2}{m^2_5}\right) \right. \right. \nonumber \\
&&- \left. \left. m^2_4 \ln \left(\frac{m^2_4}{m^2_2}\right)
\ln \left(\frac{m^2_4}{m^2_5}\right)
- m^2_5 \ln \left(\frac{m^2_5}{m^2_2}\right)
\ln \left( \frac{m^2_5}{m^2_4} \right) \right] \right. \nonumber \\
&-& \left. (m^2_4-m^2_2-m^2_6) \left[ \xi_{246}
- m^2_2\ln\left(\frac{m^2_2}{m^2_4}
\right) \ln\left(\frac{m^2_2}{m^2_6}\right) \right. \right. \nonumber \\
&&- \left. \left. m^2_4\ln\left(\frac{m^2_4}{m^2_2}\right)
\ln\left(\frac{m^2_4}{m^2_6} \right)
- m^2_6 \ln\left(\frac{m^2_6}{m^2_2}\right)\ln\left(\frac{m^2_6}{m^2_4}\right)
\! \right] \! \right]
\end{eqnarray}
where $\xi_{ijk}$ $=$ $\xi(m^2_i,m^2_j,m^2_k)$ and it is evaluated according to
eqns (A.14) or (A.17) depending on whether the particular $a^2$ is positive or
negative. A further check on our manipulations to obtain eqn (A.23) is the
absence of the arbitrary mass $\mu$ which was required at intermediate steps
to have logarithms whose arguments were dimensionless quantities.

Although it may appear that the final result is singular in certain cases
through denominator factors like $(m^2_1-m^2_2)$ when $m^2_1$ $=$ $m^2_2$, the
expression within the square brackets also vanishes. Moreover, if one sets
$m^2_2$ $=$ $m^2_1$ $+$ $\delta$, where $\delta$ is small, and expands in
powers of $\delta$ then in the limit as $\delta$ $\rightarrow$ $0$ a non-zero
non-singular function of the independent mass remains. Further, there is no
difficulty with singularities when one or more masses is zero. To illustrate
this point explicitly we consider the integral $\Lambda_{123056}$ where the
zero subscript means the corresponding mass of eqn (A.1) is zero. Its
form can readily be deduced from eqn (A.23) by taking the $m^2_4$ $\rightarrow$
$0$ limit. However, to do this the  behaviour of $\xi(x,y,z)$ in the
$z$ $\rightarrow$ $0$ limit is required since eqn (A.23) has terms like $\ln
m^2_4$ which are potentially infinite in the limit we require. It is is easy to
deduce from the explicit representation, eqn (A.14), that
\begin{equation}
\xi(x,y,z) ~ \sim ~ (x-y) \left[ 2 \mbox{Li}_2\left(1-\frac{y}{x}\right)
+ \ln \left(\frac{x}{y}\right) \ln \left(\frac{x}{z}\right)\right]
\end{equation}
as $z$ $\rightarrow$ $0$. Thus a little algebra leads to the compact
expression,
\begin{eqnarray}
\Lambda_{123056} &=& - \, \frac{1}{4(4\pi)^4(m^2_1-m^2_2)m^2_3(m^2_5-m^2_6)}
\nonumber \\
&\times& \left[(m^2_3-m^2_1-m^2_5)\xi_{135}
- (m^2_3-m^2_1-m^2_6)\xi_{136} \right. \nonumber \\
&-& \left. (m^2_3-m^2_2-m^2_5)\xi_{235}
+ (m^2_3-m^2_2-m^2_6)\xi_{236} \right. \nonumber \\
&-& \left. \rho(m^2_3,m^2_1,m^2_5) + \rho(m^2_3,m^2_1,m^2_6) \right.
\nonumber \\
&+& \left. \rho(m^2_3,m^2_2,m^2_5) - \rho(m^2_3,m^2_2,m^2_6) \right.
\nonumber \\
&+& \left. \lambda(m^2_1,m^2_5) - \lambda(m^2_1,m^2_6) - \lambda(m^2_1,m^2_5)
+ \lambda(m^2_2,m^2_6) \right]
\end{eqnarray}
with
\begin{eqnarray}
\rho(x,y,z) &=& (x-y-z) \left[ x \ln \left(\frac{x}{y}\right)
\ln\left(\frac{x}{z}\right)
+ y \ln \left(\frac{y}{x}\right) \ln\left(\frac{y}{z}\right) \right.
\nonumber \\
&& ~~~~~~~~~~~~~~~ \left. + ~ z \ln \left(\frac{z}{x}\right)
\ln\left(\frac{z}{y}\right) \right]
\end{eqnarray}
and
\begin{equation}
\lambda(x,y) ~=~ (x+y) \left[ 2(x-y) \mbox{Li}_2 \left( 1 - \frac{y}{x} \right)
- y \ln \left( \frac{x}{y} \right) \right]
\end{equation}
where $\mbox{Li}_2(t)$ is the dilogarithm function. Its properties have been
discussed extensively in \cite{J6} but we make use of the following ones here
\begin{eqnarray}
\mbox{Li}_2(-t) + \mbox{Li}_2(-1/t) &=& - \, \zeta(2) - \half \ln^2 t
\quad\quad ~~~~\mbox{for}~ t > 0 \nonumber \\
\mbox{Li}_2(t) + \mbox{Li}_2(1-t) &=& \zeta(2) - \ln t \ln (1-t)
\end{eqnarray}
and its integral representation is, \cite{J6},
\begin{equation}
\mbox{Li}_2(t) ~=~ - \, \int_0^t \frac{ds}{s} \ln (1-s)
\end{equation}
where $\mbox{Li}_2(1)$ $=$ $\zeta(2)$ $=$ $\pi^2/6$.

Finally, another check on our overall expression eqn (A.23) is the comparison
with the earlier result of \cite{BM} where only $m^2_3$ and $m^2_4$ are
non-zero, ie $\Lambda_{003400}$, which was evaluated by an independent method.
We can easily deduce an expression for $\Lambda_{003400}$ from eqn (A.25) by
using the relation (A.24) or by returning to the $I_{ijk}$ representation of
eqn (A.1) and taking the appropriate limits in that case. Useful for the former
approach are the properties of the dilogarithm function, \cite{J6}. Whilst in
the latter instance we made use of the Tyalor expansion of the $I_{ijk}$ about
zero mass and in particular,
\begin{equation}
\left. \frac{\partial^2I(x,y,z)}{\partial y \partial z} \right|_{~y~=~z~=~0}
{}~=~ \frac{\Gamma^2(\half d-2)\Gamma(4-\half d) \Gamma(5-d) x^{d-5}}
{(4\pi)^d(\mu^2)^{d-4} \Gamma(\half d)}
\end{equation}
whose $\epsilon$ expansion is easy to determine. Consequently, we find
\begin{equation}
\Lambda_{003400} ~=~ - \, \frac{1}{(4\pi)^4(m^2_3-m^2_4)} \ln \left(
\frac{m^2_3}{m^2_4} \right)
\end{equation}
This is in total agreement with the explicit calculation of \cite{BM} and is a
necessary non-trivial check that we have the overall normalization of our
integral correct, in terms of signs and factors of $2\pi$.

\vspace{1.0cm}
\noindent
{\bf Acknowledgements.} We are grateful to Naoya Hata and
and Paul Langacker for providing us with a computer-readable version of their
figure $12$ from Reference \cite{Hata}. RG would like to thank K.S. Babu
for very useful discussions on Ref. \cite{BM} and Bala Sundaram for discussions
and generous help.
 JAG thanks C. Ford and D.R.T. Jones for useful
discussions on \cite{J4} and DC thanks the Mehta Research Institute for
hospitality while this work was in progress.

\eject
\begin{center}
{\bf Figure Captions}
\end{center}
\vskip .5in
\noindent {\bf Figure $1$:} The two-loop diagram which gives rise to the
mass corrections considered in this paper.
\vskip .3in

\noindent{\bf Figure $2$:} The MSW solution space for the solar neutrino
deficit, from Ref. \cite{Hata}. Superposed on it are the $3$ curves (a),
(b) and (c) which represent sample calculations using our results. Each
curve shows the mass squared differences and mixings for the two light
neutrinos which acquire  masses radiatively, for fixed values of the masses
of the other two neutrinos which are massive at tree level.
Curve (a) in Figure $2$ corresponds to a singlet mass
of $100 $ GeV and a $\nu_{\tau}$ mass of $\approx 8.6
$ eV. Curve (b) corresponds
to a singlet mass of $ 400 $ GeV and a $ \nu_{\mu}$ mass of $7 $ eV.
 Finally, curve (c) represents a singlet mass of $1 $TeV and a $\nu_{\mu}$
mass of $\approx 9.8 $ eV.

\begin{thebibliography}{99}
\bibitem{div} A consistent use of any
regularization scheme would, of course, demonstrate that all the infinities in
eqn.(\protect\ref{GIM-1}) cancel,
thereby leaving a finite result. However there is a
simpler way to see this. Note that the mass correction is independent of the
gauge choice and instead of the unitary gauge we could have chosen to work
in the general 'tHooft--Veltman $R_\xi$ gauge instead. Of these, the Feynman
gauge has the added advantage that the gauge boson propagators are free of
the longitudinal mode. This directly reduces the naive degree of divergence
by 4. Of course, now we have to include, in addition to Fig 1, three further
diagrams with the charged Goldstones replacing one or both of the $W$'s. But
these obviously lead to terms proportional to the fermion masses.
\end{thebibliography}
\end{document}